# Two-dimensional ternary locally resonant phononic crystals with a comblike coating

Yan-Feng Wang[1], Yue-Sheng Wang[1,*], and Litian Wang[2]

[1] Institute of Engineering Mechanics, Beijing Jiaotong University, Beijing, 100044, China

[2] Department of Mechanical Engineering, Østfold University College, Halden, 1757, Norway

**Abstract**

Two-dimensional ternary locally resonant phononic crystals can be used for vibration control and noise insulation in the low (even audible) frequency range. They used to consist of cylindrical scatterers with uniform coatings in their exterior. An alternative coating scheme with a comblike profile is proposed and investigated in this paper. The band structures are calculated by using the finite element method. It is found that a complete bandgap at a significantly low frequency, the wavelength of which is more than 20 times of the lattice constant, can be induced. The mechanism for such a change is suggested by using an equivalent spring-mass model based on analyzing the eigen modes at the bandgap edges. Numerical results and the results predicted by the spring-mass model are coherent.

---

[*] Corresponding author: yswang@bjtu.edu.cn (Y. S. Wang)

## 1. Introduction

A growing interest has been focused on the periodic, elastic structures, called phononic crystals (PCs) [1]. The PCs, composed of two or more different kind of materials with different mechanical properties and mass densities, may exhibit bandgaps, within which the propagation of elastic or acoustic waves is prohibited. Two mechanisms may give rise to bandgaps, one is the Bragg scattering, and the other is the local resonance [2]. The latter can induce a complete bandgap at a frequency which is lower by two order of the magnitude than the former one. So the locally resonant PCs may have potential applications in the control of noise or vibration at low frequencies [3].

So far, various kinds of locally resonant PCs have been investigated [2-5]; and some equivalent analytical models are developed to evaluate the resonant frequencies or the lowest bandgap [6-8]. Wang *et al.* [7] examined the bandgap edge modes of the ternary locally resonant PCs, which are composed of lead spheres (solid cores) coated by silicon rubber (soft coatings) and embedded in hard epoxy (matrix). At the lower edge of the bandgap, the core oscillates as a rigid sphere and the coating acts as springs. While at the upper edge of the bandgap, the core and the matrix oscillate in a reverse phase. Based on the analysis of the edge modes, they developed equivalent spring-mass models to predict the bandgap edges. In their model, only the tensile spring was considered. The tensile spring constant was calculated from the tensile elastic constant of the coating which was approximated as a plane layer [6] between the cylinder and the matrix. Their results showed that a bandgap at a much lower frequency might be obtained by decreasing the modulus of the coating with the decrease of the bandgap width. However, it is not easy to find a material with very small modulus. Another choice is to decrease the area or volume of the coating by introducing holes.

In this paper, we propose a two-dimensional PC with a comblike coating by breaking the uniform coating into $N$ discrete pieces, see Fig. 1(a). The sectorial pieces with the central angle of $2\Delta(=\pi/N)$ are equispaced in azimuth and oriented in the radial direction. Each piece [Fig. 1(c)] can be regarded as a spring in vibration, where $\theta_i$ is the angle formed by the center line of the $i$th piece and the vibration direction of the core and is given by $\theta_i = 2\pi(i-1)/N$. The objective is to introduce free surfaces on every sectorial piece and to reduce the effective modulus of the springs, which will lead to a lower resonant frequency.

## 2. Numerical Results

The band structures and the edge modes at the bandgap edge modes of the proposed locally resonant PC with a comblike coating are calculated by using the finite element software COMSOL. The detailed process can be found in Ref. [9]. The proposed ternary locally resonant PC is composed of cylindrical metal cores coated by rubber and embedded in the polymer matrix in a square lattice. The material parameters are listed in Table I. The inner and outer radii of the coating are $r_1$ and $r_2$, respectively. And the lattice constant is $a$. The band structures for the PC with a coating of 16 pieces are shown in Fig. 2(a). For comparison, the results for the PC with a uniform coating are also presented, see Fig. 2(b). Here the reduced frequency $\Omega = \omega a/(2\pi c_t)$ (with $c_t$ being the transverse wave velocity of the matrix) is used. It is noted that a complete bandgap between the 3rd and 4th bands appears in the frequency range of $0.0282 < \Omega < 0.0572$ for the comblike coating or $0.0714 < \Omega < 0.130$ for the uniform coating, respectively. By introducing the comblike coating, the bandgap is obviously lowered as expected. The wavelength of the wave inside the bandgap is more than 20 times of the lattice constant.

The associated edge modes of the amplitude of the displacements are shown in Fig. 3. At the lower edge of the bandgap [Figs. 3(a) and 3(c)], the steel core oscillates holistically, while the matrix is still. The coating acting as springs linking the core and the matrix. At the upper edge of the bandgap [Figs. 3(b) and 3(d)], however, the matrix also oscillates, in a reverse phase to the vibration of the steel core. Thus, similar to that in Ref. [7], equivalent 'mass-spring-fixture' and 'mass-spring-mass' models can be developed to represent these vibration modes, which will be discussed in detail in the next section. For the comblike coating, each piece plays different roles in vibration: some as tensile springs, some as shear springs, and the rest as a combination of both. The shear deformation of the comblike coating reduces the effective stiffness, and thus results in a lower bandgap.

The variation of the bandgap edges with the piece number (*N*) is shown in Fig. 4. The results for the PC with a uniform coating (*N*=0) are also marked in the figure. With the increase of *N* in Fig. 4(a) ($r_1/a$=0.27), the lower edge of the bandgap first decreases quickly and then tends to a fixed value ($\Omega = 0.0258$); and the upper edge of the bandgap first decreases sharply, then increases, and finally decreases in the same manner as the lower edge of the bandgap. The bandgap nearly disappears when *N*=2, then enlarges again at *N*=3 and finally nearly unchanged ( $0.0258 < \Omega < 0.0523$ ) with *N* increasing. To understand the narrowing or disappearing of the bandgap at *N*=2, we illustrate the eigen modes of the amplitude of the displacements for PC with the coating of two pieces in Fig. 5. It is noted that the lower edge mode is similar to that of the uniform coating. While at the upper edge of the bandgap, the steel core and the matrix oscillate in a reverse phase along the direction without the coating pieces. So in this case, different from that of the lower edge, the coating acts as shear springs with small effective stiffness, which reduces the eigen frequency of the upper edge of the bandgap significantly, and hence the width of the bandgap. For comparison, we calculate the band structure of the PC with a 'virtual' uniform coating, of which the elastic modulus is a half of rubber (because the total area of the comblike coating is a half of that of the uniform coating). The two dotted lines in Fig.4 mark the lower/upper bandgap edges of the bandgap, and they are all higher than those for the comblike coating when *N*≥3. So the descent of the bandgap edges by introducing the comblike coating is not only because of the decrease of the area of the coating, but also a consequence of the shear deformation of the comblike coating. Similar trend of the changes of the bandgap edges with *N* is shown in the case with larger steel cores ($r_1/a$=0.33) , see Fig. 4(b). The most striking modification occurs when *N*=2, where a small complete bandgap exists. A thinner coating exhibits a larger effective stiffness and thus results in opening of this small bandgap. When *N*≥3, the complete bandgap tends to appear in the fixed frequency range of $0.0346 < \Omega < 0.0846$.

### 3. Equivalent spring-mass models

The observations in preceding section cannot be explained simply on the basis of the original locally resonant model [2] and the effective mass model [7] because of (a) topological difference of the coating layer and (b) introduction of free surfaces. In order to understand the mechanism of the bandgap generation and estimate the bandgap edges, we will propose an equivalent mass-spring model by taking both tensile and shear deformation, and the effect of the free surfaces into account.

We first consider the PC with a uniform coating. Following the basic idea of Ref. [7],

there exists a standing point that is immoveable in the corresponding upper edge mode. Therefore the mass of the partial coating, which is mainly compressed or stretched, was divided into two parts and added to the mass of the two oscillators. However, in this paper, the mass of the whole coating is divided into two parts; and thus the effective masses of the two oscillators are, respectively, represented by

$$m_1 = m_{core} + \alpha m_{coating}/(1+\alpha), \tag{1}$$

and

$$m_2 = m_{matrix} + m_{coating}/(1+\alpha), \tag{2}$$

where $\alpha = m_2/m_1 = (m_{matrix} + m_{coating})/(m_{core} + m_{coating})$. These equations are indeed the modification of Eq. (10) in Ref. [7].

The coating is considered as a sum of many slenders (with the central angle $d\theta$) along the $\theta$-direction, see Fig. 1(b), where $\theta$ is the angle between each slender bar and the direction of the wave propagation. Different from Ref. [7], each slender bar is regarded as a tiny spring with both tensile and shear deformation. The effective tensile stiffness of a slender bar is

$$dK_t = C_{11}\frac{r_1 d\theta}{r_2 - r_1}\cos^2\theta, \tag{3}$$

and the effective shear stiffness is

$$dK_s = C_{44}\frac{r_1 d\theta}{r_2 - r_1}\sin^2\theta, \tag{4}$$

where $C_{11} = E(1-\nu)/((1+\nu)(1-2\nu))$ and $C_{44} = E/(2(1+\nu))$ are the elastic constants for the isotropic elastic coating with $E$ and $\nu$ being the Young's modulus and Poisson's ratio. Then the total effective stiffness of the entire coating may be obtained by considering contributions of all these tiny springs. The result is

$$\bar{K} = \int_0^{2\pi}(dK_t + dK_s) = \frac{\pi r_1(C_{11} + C_{44})}{r_2 - r_1}, \tag{5}$$

where the shear constant $C_{44}$ represents the contribution of the shear deformation of the coating. Thus, the eigen frequencies can be obtained by

$$\omega_1 = \sqrt{\bar{K}/m_1}, \tag{6}$$

for the lower edge of the bandgap, and

$$\omega_2 = \sqrt{\bar{K}/m_1 + \bar{K}/m_2}, \qquad (7)$$

for the upper edge of the bandgap.

The above proposed model is different from the one developed in Ref. [7]. The latter one excludes the shear deformation of the coating. The results predicted by these two models are listed in Table II with the numbers in the brackets being the relative errors. Comparison of the numerical results from FEM and predicted results by these two models shows that both models can give satisfied predicted results for both lower and upper edges of the bandgap with small relative errors. The small mismatch between these two predicted results is mainly due to the fact that $C_{11} \gg C_{44}$. That is to say, the two models give very similar results only in this particular case.

The above proposed model is developed for the PC with a perfect coating. Then, can it be used for the PC with a comblike coating? To answer this question, we applied it directly to evaluate the lowest bandgap for the PC with a comblike coating. In this case, Eq. (5) should be rewritten as

$$\bar{K} = \sum_{i=1}^{N} \int_{\theta_i - \Delta}^{\theta_i + \Delta} (\mathrm{d}K_t + \mathrm{d}K_s). \qquad (8)$$

Particular attention should be paid on the evaluation of the upper edge for $N=2$, in which case the vibration mode is shown in Fig. 5(b) and thus $\theta_i = \pi(i-1) + \pi/2$. Evaluation of Eq. (8) with substantiation of Eqs. (3) and (4) yields

$$\bar{K} = \frac{\pi r_1 (C_{11} + C_{44})}{2(r_2 - r_1)} \qquad (9)$$

for $N \neq 2$, and

$$\bar{K} = \begin{cases} \dfrac{r_1}{r_2 - r_1}\left[\dfrac{C_{11} + C_{44}}{2}\pi - (C_{11} - C_{44})\right], & \text{upper edge of the bandgap} \\ \dfrac{r_1}{r_2 - r_1}\left[\dfrac{C_{11} + C_{44}}{2}\pi + (C_{11} - C_{44})\right], & \text{lower edge of the bandgap} \end{cases} \qquad (10)$$

for $N=2$. The difference between the lower and upper edges for $N=2$ is directly induced by the different roles of the coating in the vibration mode. It is noted that $\bar{K}$, and therefore the eigenfrequencies, are independent of $N$ except $N=2$. The predicted results from this model are shown as the thin-dashed and thin-solid lines in Fig. 4, which are unfortunately higher than the FEM results for the comblike coating.

We note that area of the free surfaces increases with $N$ increasing. Therefore, its influence on the total effective stiffness $\bar{K}$ might be the key fact to understand the variation of bandgap edges with $N$. It is known that the normal stress is zero on any free surface. Therefore, a thin 'surface layer' near the free surface is in the plane-stress state in each piece; while the central part is close to the plane-strain state. However, in the above model, tensile stiffness ($K_t$) of a slender bar is calculated through $C_{11}$ [cf. Eq. (5)] by assuming the plane-strain state in the entire piece. This leads to the overestimation of the tensile spring constant of the comblike coating and thus yields the higher eigenfrequencies of the bandgap edges. In fact, for a uniform strain field along the $r$-direction, the stress in each piece is inhomogenous along the $\theta$-direction. Therefore, this should be a two-dimensional problem. For simplicity, we will replace it by a one-dimensional problem. In this sense, an improved one-dimensional model will be developed to evaluate the total effective stiffness of the coating.

The effective shear stiffness, immune to the free surface, is still given by Eq. (4). We will develop an improved model to calculate the effective tensile spring stiffness. For simplicity, we model each sectorial piece [Fig. 1(c)] as a strip infinitely long in $z$-direction ($\varepsilon_z \equiv 0$) with a $l \times 2h$ rectangular cross-section, see Fig. 6. To determine the geometry map from a sectorial piece to a rectangular strip, we consider the following facts: The comblike coating tends to be a perfect coating when $\Delta \to \pi$, in which case the whole coating, and therefore the rectangular strip, should be in the plane-strain state [6]. That is to say, $h \to \infty$ when $\Delta \to \pi$. On the other hand, when $\Delta \to 0$, each sectorial piece is very thin, in which case we should have $h \sim O(r_1 \Delta)$. Here we suggest the following relation between $(r_1, \Delta)$ and $(l, h)$:

$$l = (r_2 - r_1), \quad h = r_1[(1 - \Delta/\pi)^{-2} - 1]. \tag{11}$$

which satisfies the above requirements. Then, the deformation of the sectorial piece along

*r*-direction can be simplified as one-dimensional tension or compression of the rectangular strip along *x*-axis.

In order to describe the inhomogeneous distribution of the stress, $\sigma_x(y)$, under a uniform strain in *x*-direction, $\varepsilon_x$, we assume that the Young's modulus is various as a function of *y*, which is denoted by $E_x(y)$ and sketched in Fig. 6. As mentioned before, the free surfaces (*y*=±*h*) are in the plane-stress state, i.e. $\sigma_y = \tau_{yx} = \tau_{yz} = 0$, which also implies that $\tau_{xy} = \tau_{zy} = 0$. The deformation along the *z*-direction is obviously uniform, therefore we have $\tau_{xz} = \tau_{zx} = 0$. Thus the stress-strain relation near the free surface is

$$\sigma_x = E'\varepsilon_x, \quad \text{(at } y=\pm h) \tag{12}$$

where $E' = E_x(\pm h) = E/(1-v^2)$. The central part far from the free surface is close to the plane-strain state. Therefore $E_x(0)$ should be between $E'$ and $C_{11}$. If the strip is very thin ($h/l \to 0$), the whole strip is in the plane-stress state, and thus $E_x(y) \equiv E'$. If the strip is very thick (i.e. $h/l \to \infty$), the central part of the whole strip is nearly in the plane-strain state with the stress-strain relation given by

$$\sigma_x = C_{11}\varepsilon_x. \tag{13}$$

These facts imply that $E_x(y)$ should satisfy the following constraint conditions:

$$\begin{cases} E_x(\pm h) = E', \\ E_x(y) = E' \text{ when } h/l \to 0, \\ E_x(y_0) = C_{11} \text{ when } h/l \to \infty, \end{cases} \tag{14}$$

where $y_0$ is an arbitrary finite value. Based on the above analysis, we interpolate $E_x(y)$ by

$$E_x(y) = E' + (C_{11} - E')(1-|y|/h)(1-e^{-h/l}). \tag{15}$$

Applying Eq. (15) to the sectorial piece in Fig. 1(c), we have

$$E_r(\theta) = E' + (C_{11} - E')(1 - \frac{(1-|\theta-\theta_i|/\pi)^{-2}-1}{(1-\Delta/\pi)^{-2}-1})(1-e^{-r_1[(1-\Delta/\pi)^{-2}-1]/l}) \tag{16}$$

in the cylindrical coordinate.

To verify the rationality of the proposed model, we calculate the strain energy of two configurations. One is a sectorial piece [Fig. 1(c)] subjected to an elongation along the *r*-direction. The strain energy is calculated by FEM. The other is the corresponding rectangular strip under the same elongation along *x*-axis. The strain energy is calculated with Eq. (15). The results for sectorial pieces with different central angles and those for their corresponding rectangular strips are illustrated in Fig. 7. An agreement is found between these two configurations, especially when the coating has more pieces. Finally, the associated effective tensile stiffness of each tiny spring is computed by

$$\mathrm{d}K'_t = E_r(\theta)\frac{r_1 \mathrm{d}\theta}{r_2 - r_1}\cos^2\theta. \tag{17}$$

Then the total effective stiffness in Eq. (8) is modified as

$$\bar{K} = \sum_{i=1}^{N}\int_{\theta_i-\Delta}^{\theta_i+\Delta}(\mathrm{d}K'_t + \mathrm{d}K_s) = \frac{\pi r_1}{2(r_2-r_1)}(E' + C_{44}) \\ + (C_{11} - E')\left(1 - e^{-r_1[(1-\Delta/\pi)^{-2}-1]/l}\right)\sum_{i=1}^{N}\int_{\theta_i-\Delta}^{\theta_i+\Delta}\left[1 - \frac{(1-|\theta-\theta_i|/\pi)^{-2}-1}{(1-\Delta/\pi)^{-2}-1}\right]\cos^2\theta\,\mathrm{d}\theta. \tag{18}$$

Unlike Eq. (9), the modified total effective stiffness varies with the piece number *N*.

By substituting Eq. (18) into Eqs. (6) and (7), we can evaluate the eigen frequencies at the bandgap edges, see Fig. 4, where the thick-dashed and thick-solid lines represent the lower and upper edges of the bandgap, respectively. The agreement between the predicted results and the numerical ones is satisfactory.

When *N* is big enough, the second part in Eq. (16) approaches to zero and we have

$$E_r(\theta) \approx E'. \tag{19}$$

Then the total effective stiffness is simplified by eliminating the second part in Eq. (18), and we get

$$\bar{K} = \frac{\pi r_1(E' + C_{44})}{2(r_2 - r_1)}, \tag{20}$$

which is independent of *N*. Thus the eigenfrequencies of the bandgap edges approach constants or 'fixed values' when $N \to \infty$ as mentioned before.

Moreover, the numerical results for the PCs in a triangular lattice are also plotted in Fig.4. The steel core, the coating and their filling ratio are exactly the same as those in the square lattice. The results confirm that the locally resonant gap is invariant with respect to the lattice symmetries [2]. In other words, the proposed model can evaluate the lowest bandgap for two-dimensional locally resonant PC with a comblike coating regardless of the lattice symmetry.

Another way to develop an equivalent spring-mass model is to model the comblike coating as an 'effective homogeneous perfect coating' of which the effective bulk and shear moduli [10] are obtained based on the equivalence of the strain energy.

As indicated before, the shear modulus is uniform in a sectorial piece, immune to the free surfaces. Therefore the effective shear modulus is simply given by

$$\bar{\mu} = C_{44}/2. \tag{21}$$

The derivation of the effective bulk modulus, $\bar{k}$, is cumbersome. We present the details in Appendix. The final result is

$$\bar{k} = \frac{Nh(r_2+r_1)}{4\pi r_2^2}[E' + (C_{11}-E')(1-e^{-h/l})/2] + \frac{r_2^2-r_1^2}{2r_2^2}\bar{\mu}, \tag{22}$$

with $l$ and $h$ given by Eq. (11).

The effective bulk modulus, normalized by the shear modulus of the coating, is shown in Fig. 8. As we can see, it decreases monotonously as the number of pieces increasing, as a main consequence of the decrease of the Young's modulus in Eq. (15). Meanwhile, replacing $C_{11}$ with $(\bar{k}+\bar{\mu})$ and $C_{44}$ with $\bar{\mu}$ in Eq. (5), Eqs. (6) and (7) can be used to evaluate the bandgap edges, see Table III. The numbers in the brackets show the relative errors compared to the FEM results. Predicted results with these effective moduli by the model in Ref. [7], as well as those predicted by the modulus function in Eq. (16) are also presented for comparison. For a small piece number ($N$=4), the pieces distribute so sparsely that the comblike coating cannot be approximated as a 'perfect coating'. Then a big relative error is shown. For a big piece number ($N$=32), the pieces are evenly distribute, and we have a small relative error. Unfortunately, the bandgap edges evaluated by either the model of Eq. (5) or the one in Ref. [7] are still much underestimated. However the model of Eq. (15) or (16) can yield more accurate results.

## 4. Conclusions

In summary, when a comblike coating is introduced to the ternary locally resonant PC, a complete bandgap at a lower frequency can be obtained, except for the case of a comblike coating with two pieces. This is due to the fact that the thin layers of the 'plane-stress' state near the free surface can decrease the effective stiffness of the coating. The width of this low bandgap, which is a little smaller than that for a uniform coating, is nearly unchanged with its center frequency getting slightly low when the piece number increases. An equivalent spring-mass model is proposed for a better understanding of the mechanism of the bandgap generation as well as accurate evaluation of the corresponding bandgap edges. The research is relevant to the design of locally resonant PCs which have the potential applications in wave or noise control in the low (even audible) frequency range.

It is worthy to point out that this paper just provides a basic idea for the design of the locally resonant PCs with a comblike coating. Actually, a steel core coated with the comblike coating can be manufactured by wrapping the core with a thin layer of coating with many discrete pieces. The discrete piece can also be in other shape, such as parallelogram, trapezoid, etc. Distinguished features maybe expected with various comblike coatings.

**Acknowledgements**

The authors are grateful for the support from the National Natural Science Foundation of China (Grant No. 11272041), the National Basic Research Program of China (2010CB732104) and the Oslofjord Research Fond of Norway (RFFOFJOR-ES468298). The first author also acknowledges Rino Nilsen for fruitful discussion during his short visit at Fredrikstad.

**Appendix**

In this appendix, we show the process to calculate the effective bulk modus of the comblike coating based on the energy method. We consider the equivalence of the train energy between two states: one is a comblike coating with $N$ pieces subjected to an elongation along the $r$-direction; the other is the corresponding 'effective homogeneous perfect coating' with the effective bulk modulus $\bar{k}$ and the effective shear modulus $\bar{\mu}$ under the same elongation along the $r$-direction. The effective shear modulus $\bar{\mu}$ is given in Eq. (21). Next we determine the effective bulk modulus $\bar{k}$.

Suppose that the comblike coating is sustained an elongation $\delta$ along the $r$-direction. Its strain energy can be calculated by the sum of the strain energy of $N$ identical $l \times 2h$ rectangular strips [Fig. 6] under the elongation of $\delta$ along the $x$-axis. The strain energy density in each strip is

$$w_1 = \int_0^{\varepsilon_x} \sigma_x d\varepsilon_x = \int_0^{\varepsilon_x} E_x(y)\varepsilon_x d\varepsilon_x = \frac{\delta^2}{2l^2} E_x(y). \tag{A1}$$

Then the strain energy of the comblike coating is

$$W_1 = N \int_0^l \int_{-h}^h w_1 dx dy = \frac{Nh\delta^2}{l}[E' + (C_{11} - E')(1 - e^{-h/l})/2]. \tag{A2}$$

Deformation of an 'effective homogeneous perfect coating' under the elongation $\delta$ along the $r$-direction is an axisymmetric elastic problem of which the displacement field can be generally written as $u_r = D_1 r + D_2/r$ [11] where $D_1$ and $D_2$ are determined by the boundary condition. Suppose that the inner boundary of the coating is fixed, and the outer boundary is subjected to a displacement $\delta$. We can have $D_1 = \delta r_2/(r_2^2 - r_1^2)$ and $D_2 = -D_1 r_1^2$. Then the strain energy density in the 'effective homogeneous perfect coating' is

$$w_2 = \int_0^{\varepsilon_r} \sigma_r d\varepsilon_r + \int_0^{\varepsilon_\theta} \sigma_\theta d\varepsilon_\theta = \frac{\delta^2 r_2^2}{(r_2^2 - r_1^2)^2}[4\bar{k} + 2\bar{\mu}(r_1^4/r^4 - 1)]. \tag{A3}$$

And the total strain energy is

$$W_2 = \int_0^{2\pi} \int_{r_1}^{r_2} w_2 r dr d\theta = \frac{\pi \delta^2 r_2^2}{r_2^2 - r_1^2}[4\bar{k} - 2\bar{\mu}(1 - r_1^2/r_2^2)]. \tag{A4}$$

Equating the strain energy $W_1$ and $W_2$, we get the effective bulk modulus as

$$\bar{k} = \frac{Nh(r_2 + r_1)}{4\pi r_2^2}[E' + (C_{11} - E')(1 - e^{-h/l})/2] + \frac{r_2^2 - r_1^2}{2r_2^2}\bar{\mu}. \tag{A5}$$

**Table captions**

Table I. Elastic parameters of the materials.

Table II. Normalized Bandgap edges ($\Omega = \omega a/(2\pi c_t)$) predicted by equivalent models for PC with a uniform coating under different geometrical parameters.

Table III. Normalized Bandgap edges ($\Omega = \omega a/(2\pi c_t)$) for PC with a comblike coating evaluated by using the effective bulk and shear moludi ($r_1/a$=0.27 and $r_2/a$=0.4).

**Figure captions**

Fig.1 Configurations of the ternary PC with a comblike coating (a) and a uniform coating (b). The white part in panel (b) is vacuum. Panel (c) shows a sectorial piece of the comblike coating in panel (a).

Fig.2 Band structures of the PC with (a) a comblike coating of 16 pieces and (b) a uniform coating in a square lattice, where $r_1/a$=0.27 and $r_2/a$=0.4.

Fig. 3 Vibration modes of the amplitude of the displacements at the bandgap edges marked in Fig. 2. Panels (a)-(d) correspond to points S1-S4, respectively.

Fig.4 Variations of bandgap edges with the number of the comblike coating for $r_2/a$=0.4 and $r_1/a$=0.27 (a) or $r_1/a$=0.33 (b). The scattered symbols represent the numerical results; and the solid and dashed lines represent the analytical results. The two dotted lines represent the results for PC with a uniform coating, the modulus of which is a half of rubber. $\square$($\triangle$) — Lower edge for a square (triangular) lattice; $\blacksquare$($\blacktriangle$) — Upper edge for a square (triangular) lattice. $\mathbf{-\ -}$ ($----$) — Lower edge evaluated by using $E_r(\theta)$ ($C_{11}$); ———— (———) — Upper edge evaluated by using $E_r(\theta)$ ($C_{11}$).

Fig.5 Vibration modes of the amplitude of the displacements at the lower edge (a) and the upper edge (b) for PC with a coating of two pieces.

Fig.6 Explanation of the calculation method of parameters in the proposed model for a $l \times 2h$ rectangular strip with $l = (r_2 - r_1)$ and $h = r_1[(1-\Delta/\pi)^{-2} - 1]$, which corresponds to the sectorial piece in Fig. 1(c)

Fig.7 Strain energy versus the central angle for one sectorial piece and its corresponding rectangular strip under the same elongation.

Fig.8 Variations of the normalized effective bulk modulus with the piece number of the comblike coating for $r_2/a$=0.4.

Table I. Elastic parameters of the materials.

| Materials | Core | Coating | Matrix |
|---|---|---|---|
| Mass density, $\rho$ [kg/m$^3$] | 8950 | 1020 | 1200 |
| Young's modulus, $E$ [Pa] | $2.1\times10^{11}$ | $1\times10^{5}$ | $3.5\times10^{7}$ |
| Poisson's ratio, $\nu$ | 0.29 | 0.47 | 0.49 |

Table II. Normalized Bandgap edges ($\Omega = \omega a/(2\pi c_t)$) predicted by equivalent models for PC with a uniform coating under different geometrical parameters.

| Geometry | $r_1/a$=0.27 and $r_2/a$=0.4 | | $r_1/a$=0.33 and $r_2/a$=0.4 | |
|---|---|---|---|---|
| Bandgap edges | Lower edge | Upper edge | Lower edge | Upper edge |
| FEM results | 0.0714 | 0.130 | 0.0920 | 0.212 |
| Present model [Eq. (5)] | 0.0705 (1.3%) | 0.134 (3.1%) | 0.0905 (1.6%) | 0.209 (1.4%) |
| Model in Ref. [7] | 0.0725 (1.5%) | 0.135 (3.8%) | 0.0907 (1.4%) | 0.209 (1.4%) |

Table III. Normalized Bandgap edges ($\Omega = \omega a/(2\pi c_t)$) for PC with a comblike coating evaluated by using the effective bulk and shear moludi ($r_1/a$=0.27 and $r_2/a$=0.4).

| Piece number | N=4 | | N=32 | |
| --- | --- | --- | --- | --- |
| Bandgap edges | Lower edge | Upper edge | Lower edge | Upper edge |
| FEM results | 0.0342 | 0.0686 | 0.0267 | 0.0542 |
| Present model [Eq. (16)] | 0.0309(10%) | 0.0611(11%) | 0.0264(1%) | 0.0520(4%) |
| Present model [Eq. (5)]* | 0.0227(34%) | 0.0448(35%) | 0.0203(24%) | 0.0401(26%) |
| Model in Ref. [7] | 0.0204(40%) | 0.0403(41%) | 0.0175(34%) | 0.0345(36%) |

*Note: $C_{11}$ and $C_{44}$ in Eq. (5) are replaced by $(\bar{k}+\bar{\mu})$ and $\bar{\mu}$, respectively.

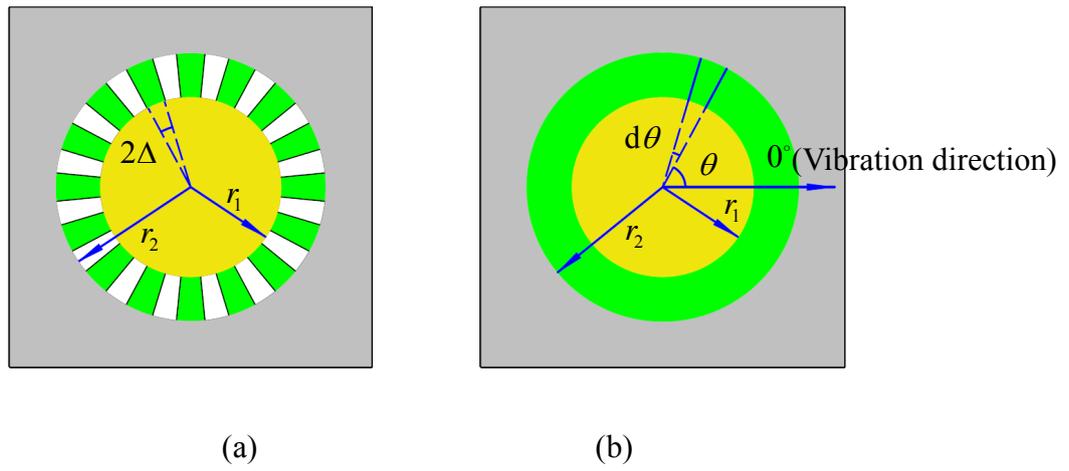

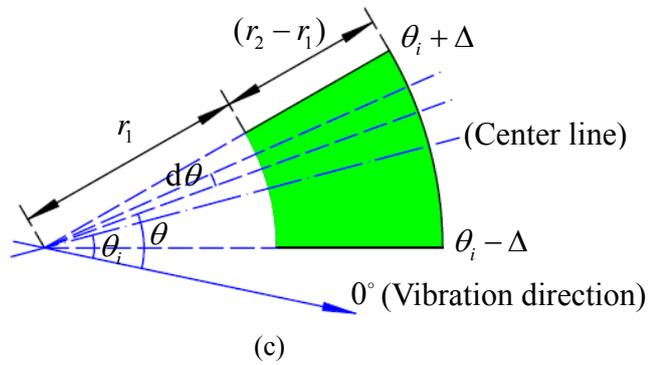

Fig.1 Configurations of the ternary PC with a comblike coating (a) and a uniform coating (b). The white part in panel (b) is vacuum. Panel (c) shows a sectorial piece of the comblike coating in panel (a).

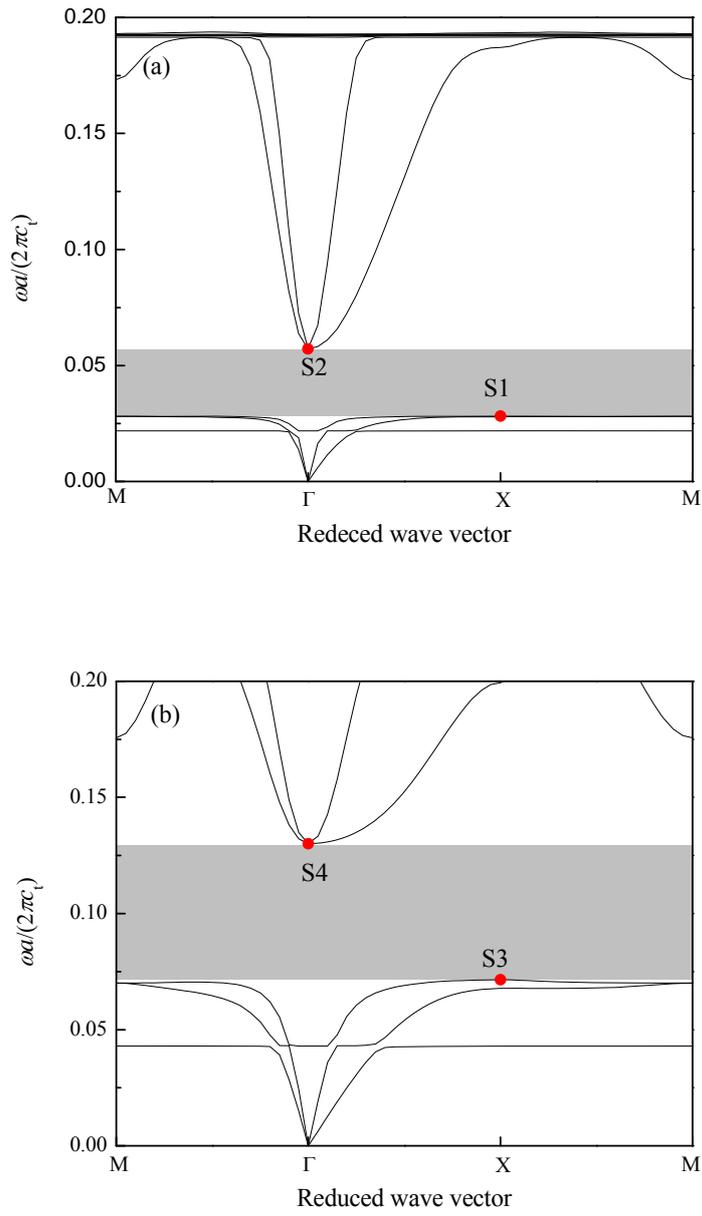

Fig.2 Band structures of the PC with (a) a comblike coating of 16 pieces and (b) a uniform coating in a square lattice, where $r_1/a$=0.27 and $r_2/a$=0.4.

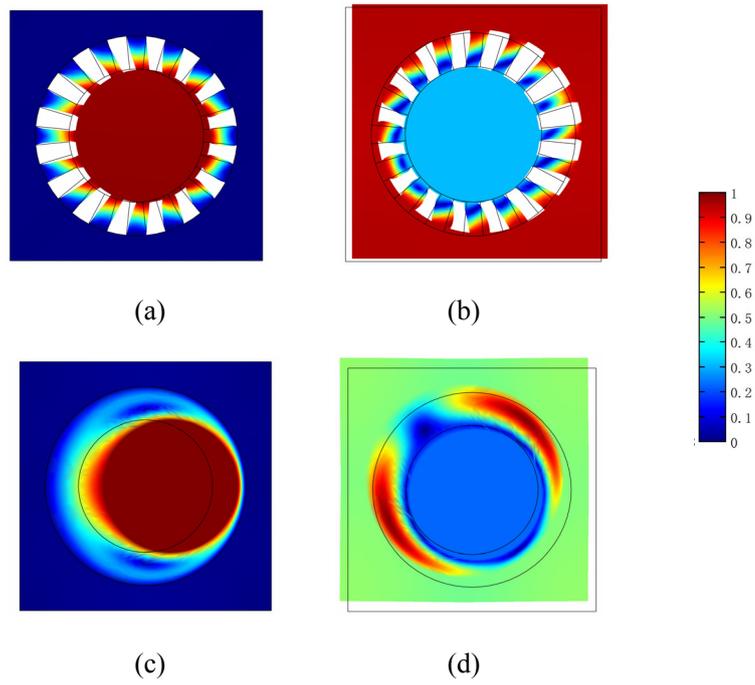

Fig. 3 Vibration modes of the amplitude of the displacements at the bandgap edges marked in Fig. 2. Panels (a)-(d) correspond to points S1-S4, respectively.

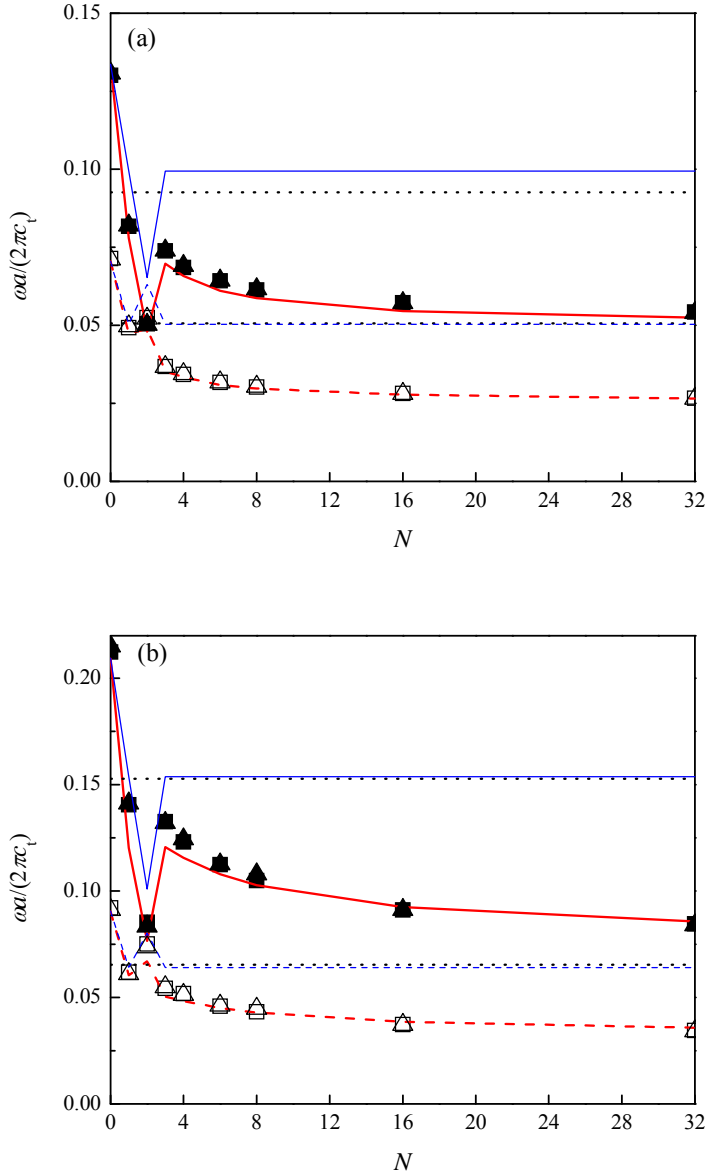

Fig.4 Variations of bandgap edges with the piece number of the comblike coating for $r_2/a=0.4$ and (a) $r_1/a=0.27$ or (b) $r_1/a=0.33$. The scattered symbols represent the numerical results; and the solid and dashed lines represent the analytical results. The two dotted lines represent the results for PC with a uniform coating, the modulus of which is a half of rubber. $\square(\triangle)$ — Lower edge for a square (triangular) lattice; ■(▲) — Upper edge for a square (triangular) lattice. ▬ ▬ (▪ ▪ ▪ ▪) — Lower edge evaluated by using $E_r(\theta)$ ($C_{11}$); ——— (———) — Upper edge evaluated by using $E_r(\theta)$ ($C_{11}$).

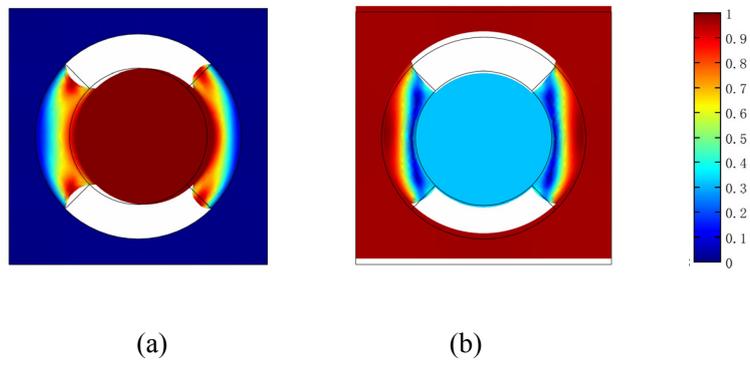

(a)　　　　　　　　(b)

Fig.5 Vibration modes of the amplitude of the displacements at the lower edge (a) and the upper edge (b) for PC with a coating of two pieces.

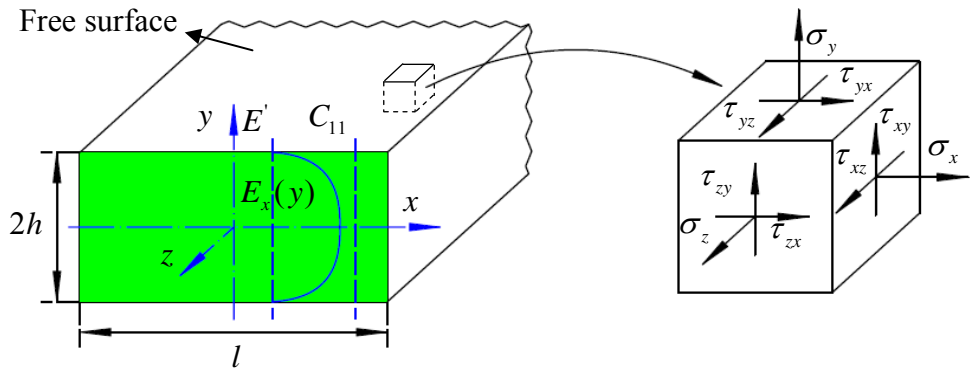

Fig.6 Explanation of the calculation method of parameters in the proposed model for a $l \times 2h$ rectangular strip with $l = (r_2 - r_1)$ and $h = r_1[(1-\Delta/\pi)^{-2} - 1]$, which corresponds to the sectorial piece in Fig.1(c).

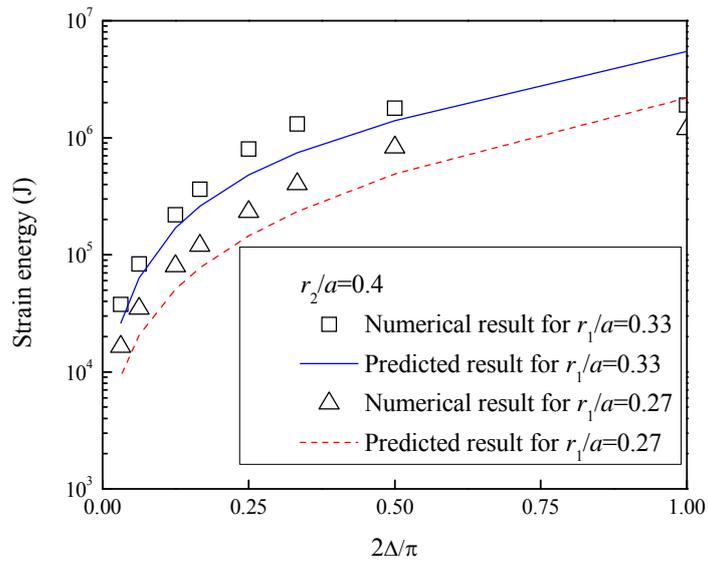

Fig.7 Strain energy versus the central angle for one sectorial piece and its corresponding rectangular strip under the same elongation.

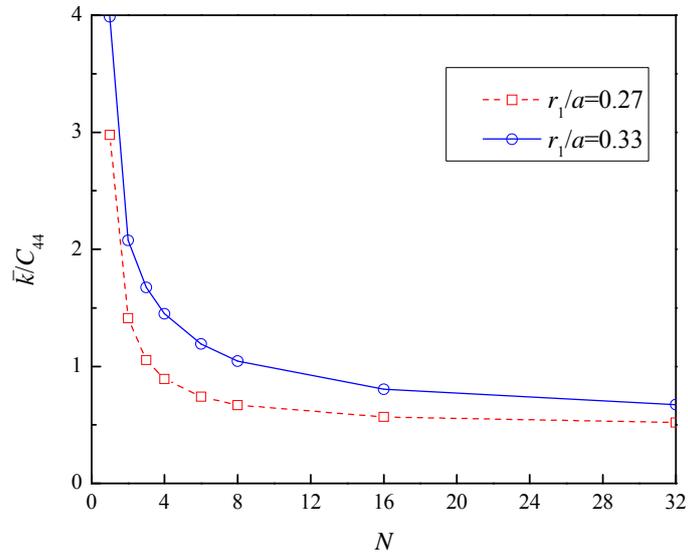

Fig.8 Variations of the normalized effective bulk modulus with the piece number of the comblike coating for $r_2/a$=0.4.